\documentclass[a4paper,final]{appolb}
\usepackage{epsfig}

\begin{document}
\title{New bounds on the mass of a b' quark}
\author{ S. M. Oliveira
\address{Centro de F\'\i sica Te\'orica e Computacional,
   Faculdade de Ci\^encias, Universidade de Lisboa,
   Av. Prof. Gama Pinto 2, 1649-003 Lisboa, Portugal}
\and
R. Santos
\address{Centro de F\'\i sica Te\'orica e Computacional,
   Faculdade de Ci\^encias, Universidade de Lisboa,
   Av. Prof. Gama Pinto 2, 1649-003 Lisboa, Portugal \\ and\\
 Instituto Superior de Transportes e
   Comunica\c{c}\~oes, Campus Universit\'ario
   R. D. Afonso Henriques, 2330-519 Entroncamento, Portugal}}

\maketitle
\begin{abstract}
In this work we present limits on a sequential down-type
quark, $b'$, based on the most recent DELPHI data. Using
all available experimental data for $m_{b'} > 96$ GeV
we conclude that a sequential four generations model
is far from being experimentally excluded.
\end{abstract}


\section{Introduction}

The Standard Model (SM) of the electroweak interactions
start to show some difficulties in explaining recent
experimental data especially in the fermionic sector.
What about a fourth family of fermions? Is is experimentally
ruled out?
Cancellation of gauge anomalies require the addition
of a family of leptons for each family of quarks added
to the SM. Since
the number of light neutrinos ($ m_{\nu} \, < \, M_Z/2$)
is definitely equal to three \cite{Hagiwara:fs} the new
lepton family has to accommodate a neutrino
with a mass larger than around 45 GeV. Hence, if a sequential
fourth family exists it certainly has to show
a much different
structure in the leptonic sector.

Despite the strength of the previous argument
one should try to experimentally exclude the existence
of a fourth generation. In fact such evidence does
not yet exist. 
The most recent precision electroweak results
\cite{Novikov:2001md}
allow a sequential
fourth generation if the quark masses
are not too far apart.
The same results also disfavour
a degenerate fourth family if both the leptonic and hadronic
sector are degenerate. This is in agreement
with the conclusions of Erler and Langacker
\cite{Hagiwara:fs}. However, it was shown in 
\cite{Frampton:1999xi} that even if one takes a degenerate
fourth family of quarks with 150 GeV masses, it is enough
to choose a non-degenerate family of leptons
with masses of 100 GeV and 200 GeV and a Higgs
mass of 180 GeV for the discrepancy with experimental data
to fall from roughly
three to two standard deviations.  
Moreover, it is clear
that any new physics
will also influence these results.

It was shown in refs. \cite{Frampton:1999xi, Arhrib:2000ct}
that the mass range $|m_{t'}-m_{b'}| \leq 60$ GeV, where
$t'$ and $b'$ are the fourth generation quarks,
is consistent with the precision electroweak data
on the $\rho$ parameter.
This range enable us to say that even if 
$m_{b'}>m_{t'}$, the decay $b' \rightarrow t' \, W$
is forbidden. The decay $b' \rightarrow t' \, W^*$
although allowed, is phase space suppressed
and
consequently extremely small in the mass range under study
(from now on we consider $m_{b'} < m_{t'}$).
Experimental data allow us 
to go only up to $m_{b'}$ close to $190$ GeV. Hence,
the $b'$ can not decay to a top quark.
Furthermore,
while some recent studies (see \cite{us})
have constrained the Cabibbo-Kobayashi-Maskawa (CKM) elements
of the fourth generation, they
do not influence our results. Nevertheless we will take
into account the $2 \sigma$ bound 
$|V_{tb}|^2 + 0.75 |V_{t'b}|^2 \leq 1.14$  \cite{Yanir:2002cq}
coming
from $Z \rightarrow b \bar{b}$ to constrain the CKM 
element $V_{cb'}$.

There are presently three bounds on the $b'$ mass
for $m_{b'}\, > \, 96$ GeV and all of them
suffer from the drawback of assuming a
100 \% branching ratio for a specific
decay channel.
The first one \cite{Affolder:1999bs}, $m_{b'}\, > \, 199$ GeV, assumes
that $Br(b' \rightarrow b \, Z)=100 \%$. 
We will drop this condition and use instead their plot of
$\sigma (p \, \bar{p} \rightarrow b' \bar{b'} + X) \times
Br^2(b' \rightarrow b \, Z)$ as a function of the $b'$ mass.
The second one \cite{Abachi:1995ms} $m_{b'}\, > \, 128$ GeV, is 
based on the data collected in the top quark search. Because
the D0 collaboration looked for $t \rightarrow b \, W$, the analysis
can be used to set a limit on 
$\sigma (p \, \bar{p} \rightarrow b' \bar{b'} + X) \times
Br^2(b' \rightarrow c \, W)$. By doing so we assume that the
$b$ and $c$ quark masses are negligible and that
$\sigma (p \, \bar{p} \rightarrow b' \bar{b'}) \approx
\sigma (p \, \bar{p} \rightarrow t \bar{t})$. The obtained
limit
$m_{b'}\, > \, 128$ GeV assumes
$Br(b' \rightarrow c \, W) = 100 \%$.
The third bound is from CDF \cite{Abe:1998ee} and is based on
the decay $b' \rightarrow b \, Z$ followed by the search for
$Z \rightarrow e^+ \, e^-$ with displaced
vertices.
They also assume $Br(b' \rightarrow b \, Z)=100 \%$
and their excluded region depends heavily on the $b'$
lifetime. This means we can not use this limit
in our analysis (for a discussion see \cite{us}).

Hence, we think it is worthwhile to reexamine the
limits on the $b'$ mass. We will use
the CDF and the D0 data which, together with
the new DELPHI data \cite{DELPHI_NOTE}, is all there is  available
for $m_{b'} > 96$ GeV.
We will draw exclusion plots in the planes
($R_{CKM}$, $m_{b'}$) and ($m_{t'}$, $m_{b'}$), where
$R_{CKM}=|\frac{V_{cb'}}{V_{tb'} \, V_{tb}}|$,
from 96 GeV
to 190 GeV without assuming a definite value for the
branching ratios of specific channels.
Notice that the use of the $R_{CKM}$ 
variable provides a new way to look at the experimental results.
This variable
enable us to actually use and combine all the available
data. Moreover, the new form in which the results are 
presented
will serve as a guide to future experiments since
it is possible to know how far one has to go
to exclude the regions that are still allowed.

\section{\lowercase{b'} production and decay}

At LEP, a pair of  $b' \, \bar{b'}$ quarks is produced
via $e^+ \, e^- \rightarrow b' \, \bar{b'}$.
The corresponding cross section was calculated 
using \textsc{Pythia} \cite{Sjostrand:2000wi}
with initial state radiation (ISR), final state radiation
and QCD corrections turned on.
The results were cross checked in \cite{us} and agree
very well with the \textsc{Pythia} results.

The equivalent process at the Tevatron is
$p \, \bar{p} \rightarrow b' \, \bar{b'} + X$
with the relevant processes being
$gg \, (q \, \bar{q}) \rightarrow b' \, \bar{b'}$.
Due to its hadronic nature, this cross section
is equal to the top quark production
one and it is known to order
$\alpha_s^3$ \cite{Laenen:1993xr}.
This approximation is used both by the CDF
and the D0 collaborations in their studies
on $b'$ production and decay.

Two body $b'$ decays occur either through neutral currents
(NC)
or through charged currents (CC). Although NC
proceed only via loops, it was shown in
\cite{Hou:1988yu} that depending on the values
of the CKM matrix elements and on the values of
the quark masses, they can be comparable to 
CC decays. The reason is simple:
if $b' \rightarrow W \, t$ and 
$b' \rightarrow W \, t'$ are not allowed,
the dominant CC decay is 
$b' \rightarrow W \, c$ which is doubly
Cabbibbo suppressed.

As long as the Higgs channel is closed
the dominant neutral decay is
$b' \rightarrow b \, Z$. Other neutral decays like 
$b' \rightarrow b \, g$ and
$b' \rightarrow b \, \gamma$ 
give smaller
contributions but can sometimes be relevant.
As soon as the Higgs channel
opens the decay $b' \rightarrow b \, H$ can be as large
as $b' \rightarrow b \, Z$.

The three body decays
$b' \rightarrow b \, e^+ \, e^-$,
$b' \rightarrow b \, \nu \, \bar{\nu}$ and 
$b' \rightarrow b \, q \, \bar{q}$, including box diagrams
were calculated in \cite{Hou:1989ty}. Using the
results in \cite{Hou:1989ty} we estimate \cite{us}
that
all three body decays plus the decay
$b' \rightarrow b \, \gamma$
are smaller than
$b' \rightarrow b \, g$. Nevertheless,
because we want to make a conservative estimate we
will take it to be as large as
$b' \rightarrow b \, g$.

Taking $V_{ub'} \, V_{ub} \approx \, 0$
and $V_{cb} \approx \, 10^{-2}$ and
using the unitarity of the CKM matrix
we obtain  $V_{t'b'} \, V_{t'b} \approx 
- V_{tb} \, V_{tb'}$. Thus,
we can write all branching fractions
as a function of three quantities alone:
$R_{CKM}$, $m_{t'}$ and
$m_{b'}$. 

One-loop calculations of the NC $b'$ decays
were performed using the FeynArts, FeynCalc \cite{feyn}
and the LoopTools/FF \cite{ff} packages.
We have carried out several checks in the four generations
model following
\cite{Arhrib:2000ct,Hou:1988yu,Hou:1989ty}
and we have found full agreement.

\begin{figure}[htbp]
  \begin{center}
    \epsfig{file=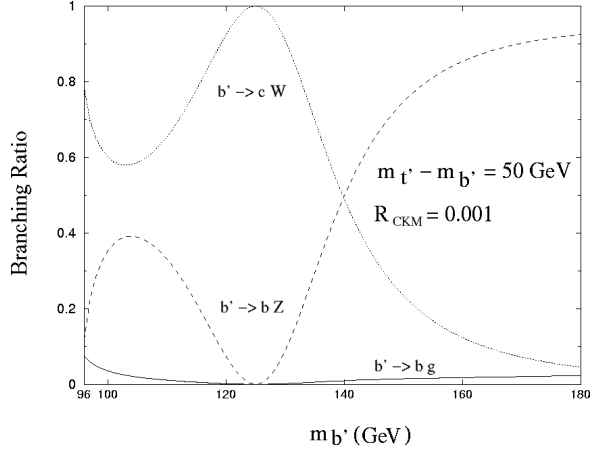,width=7.7 cm,angle=0}
    \caption{Branching ratios as a function of the $b'$ mass.
            The Higgs channel is closed.
            The dashed line is
            $b'\rightarrow \, b \, Z$; the full line is
            $b'\rightarrow \, b \, g$ and the dotted line is
            $b'\rightarrow \, c \, W$.}
    \label{fig_new1}
  \end{center}
\end{figure}
To decide on the relevant 
values of $R_{CKM}$ and $m_{t'}$ to use, we present
in fig. \ref{fig_new1} the branching ratios
as a function of the $b'$ mass with $R_{CKM}=0.001$
and  $m_{t'} - m_{b'}= 50$ GeV.
The closer
to $m_{b'}=96$ GeV we are the larger
$b'\rightarrow \, b \, g$ gets due to phase space suppression
of the competing NC  $b'\rightarrow \, b \, Z$.
In fact, for an almost degenerate fourth family and 
small values of $R_{CKM}$,
$b'\rightarrow \, b \, g$ can be the dominant
NC for  $m_{b'}=96$ GeV.
As soon as one moves away from this value,
$b'\rightarrow \, b \, Z$ becomes the dominant NC. If the Higgs
channel is closed , for $m_{b'} \geq 97$ GeV, the competition
is always between $b'\rightarrow \, c \, W$
and $b'\rightarrow \, b \, Z$. As $m_{b'}$ rises so
does the NC except if the 
Glashow-Iliopoulos-Maiani (GIM) mechanism gets in the way.
It can be clearly seen in the figure the GIM mechanism
acting for
$m_{b'} \approx 125$ GeV, that is, $m_{t'}-m_{t}=0$.
Then the NC rises again and the CC falls crossing at 140 GeV.
When $R_{CKM}$ grows so does  $b'\rightarrow \, c \, W$ and the
crossing point is shifted to the left. As the mass difference
tends to zero the GIM effect is shifted to  $m_{b'} \approx m_{t}$.

\section{Results and discussion}

Using the latest experimental data from the 
DELPHI collaboration
and the data from the CDF and D0 collaborations
together with the theoretical values of the cross
sections and the branching ratios we have drawn
the exclusion plots shown in the figures below.

\begin{figure}[htbp]
  \begin{center}
    \epsfig{file=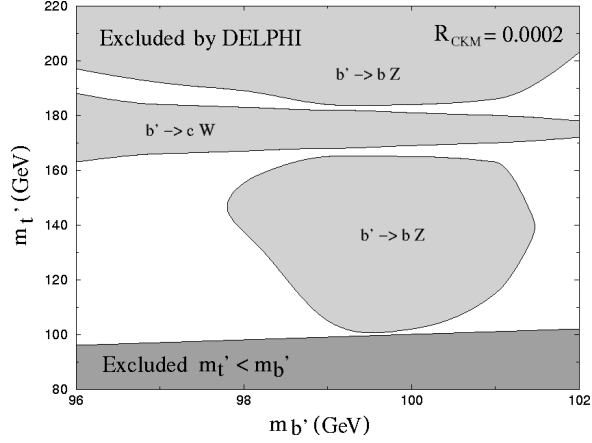,width=7.7 cm,angle=0}
    \caption{95 \% confidence level (CL) excluded region in the plane
            ($m_{t'}$, $m_{b'}$) with
            $R_{CKM}= 0.0002$,
            obtained from limits on
            $Br_{b'\rightarrow \, b \, Z}$
            and 
            $Br_{b'\rightarrow \, c \, W}$.}
    \label{fig_lip2}
  \end{center}
\end{figure}

\begin{figure}[htbp]
  \begin{center}
    \epsfig{file=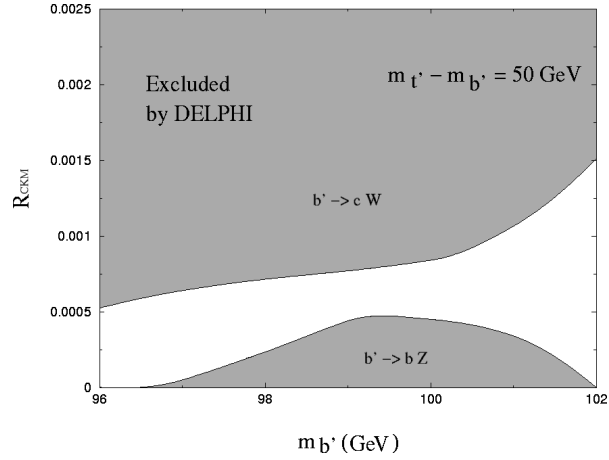,width=7.9 cm,angle=0}
     \caption{95 \% CL excluded region in the plane
            ($R_{CKM}$, $m_{b'}$) with
            $m_{t'}-m_{b'} = 50$ GeV,
            obtained from limits on
            $Br_{b'\rightarrow \, b \, Z}$ (bottom)
            and 
            $Br_{b'\rightarrow \, c \, W}$ (top).}
    \label{fig_lip1}
  \end{center}
\end{figure}

The results based on the DELPHI data,
are shown in figs. 2 and 3. 
The excluded regions in the two plots
due to the limits on $Br_{b'\rightarrow \, c \, W}$
are the stripe centred in $m_{t'}$ in fig.2 and
the upper excluded region in fig.3.
The remaining excluded regions
are due to limits on  $Br_{b'\rightarrow \, b \, Z}$.
When $(m_{t'}-m_{t}) \rightarrow 0$,
$Br_{b'\rightarrow \, b \, Z}$ decreases as a
consequence of a GIM suppression and 
$Br_{b'\rightarrow \, c \, W}$ becomes dominant.
In fact, when $m_{t'}-m_{t}=0$,  
$Br_{b'\rightarrow \, c \, W} \approx 100 \%$. Thus,
there is always an excluded stripe around  $m_{t}$.
As $R_{CKM}$ grows, \textit{i.e.}, CC dominates, the stripe gets
larger and the other two regions in fig. 2 get smaller.
This can also be seen in fig. 3 where for 
$R_{CKM}>0.0015$ everything is excluded.
Obviously this will change if we change
the mass difference.

The reason why there isn't a lower bound 
close to 96 GeV in fig.3 
is because of the competing NC. Close to
the $Z \, b$ threshold ($\approx$ 96 GeV),
$b' \rightarrow b \, g$ dominates over $b' \rightarrow b \, Z$
and the experimental bound on $Br_{b' \rightarrow b \, Z}$
becomes useless. As one moves away from the $Z \, b$ threshold,
$b' \rightarrow b \, Z$ becomes the dominant NC. 
After 102 GeV almost all values are allowed because
the experiments are not sensitive to those mass values. 

\begin{figure}[htbp]
  \begin{center}
    \epsfig{file=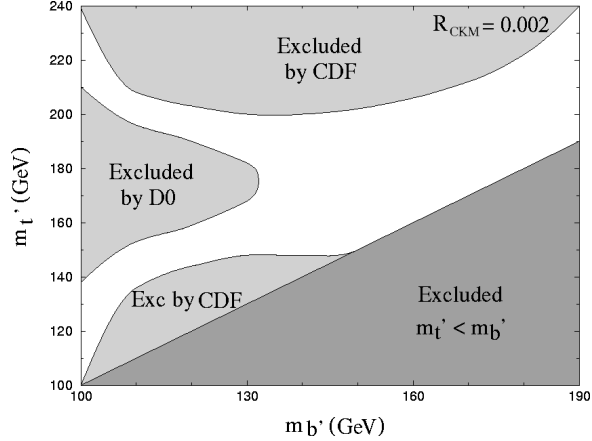,width=7.7 cm,angle=0}
    \caption{95 \% CL excluded region in the plane
            ($m_{t'}$, $m_{b'}$) with
            $R_{CKM} = 0.002$,
            obtained from limits on
            $Br_{b'\rightarrow \, b \, Z}$ by the CDF coll.
            and 
            $Br_{b'\rightarrow \, c \, W}$ by the D0 coll.}
    \label{fig_cdf2}
  \end{center}
\end{figure}

\begin{figure}[htbp]
  \begin{center}
    \epsfig{file=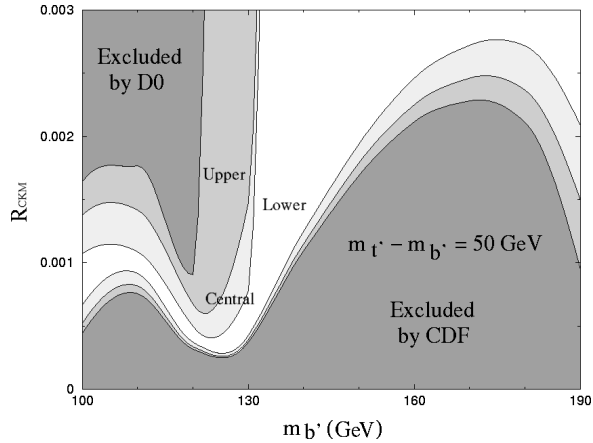,width=7.7 cm,angle=0}
     \caption{95 \% CL excluded region in the plane
            ($R_{CKM}$, $m_{b'}$) with
            $m_{t'}-m_{b'} = 50$ GeV,
            obtained from limits on
            $Br_{b'\rightarrow \, b \, Z}$ by the CDF coll. (bottom)
            and 
            $Br_{b'\rightarrow \, c \, W}$ by the D0 coll. (top).
            Upper, Central and Lower curves correspond to the values
            used for the $b'$ production cross-section.}
    \label{fig_cdf1}
  \end{center}
\end{figure}

In figs. 4 and 5 we show similar plots but using the CDF
and the D0 data. The behaviour follow the general trend explained
for the DELPHI data. The D0 deals with the CC and the CDF
deals with the NC. The three curves marked upper,
central and lower in fig. 5 are related with the theoretical
error bars in the $b'$ production cross section. In fig. 4
we have used central values for the cross sections.
Again and for the same reason we see a stripe around
$m_{t}$ in fig. 4. The stripe ends, as it should,
for $m_{b'}$ close to 130 GeV
which is approximately the D0 bound on $m_{b'}$. 
Notice that whatever the value of  $m_{b'}$ is, one can
always find an allowed $m_{t'}$ if $R_{CKM}$ is not too
large. As $R_{CKM} \rightarrow 0$,
$Br_{b' \rightarrow b \, Z} \approx 100 \%$ and we
recover the CDF bound \cite{Affolder:1999bs}.


In some cases the allowed regions in the CDF/D0 and DELPHI
plots overlap and the excluded region grows. For instance,
considering $m_{b'} = 100$ GeV 
and $m_{t'}-m_{b'} = 50$ GeV we get for DELPHI
$4.5 \times 10^{-4} < R_{CKM} \, < \, 8.4 \times 10^{-4}$
and for CDF/D0 (lower)
$6.7 \times 10^{-4} < R_{CKM} \, < \, 1.1 \times 10^{-3}$.
Hence, the resulting excluded region is
$6.7 \times 10^{-4} < R_{CKM} \, < \, 8.4 \times 10^{-4}$.

With the bound 
$|V_{tb}|^2 + 0.75 |V_{t'b}|^2 \leq 1.14$  \cite{Yanir:2002cq}
and assuming $|V_{tb}| \approx 1$ it is possible
to limit the value of the matrix element 
$V_{cb'}$. For the same value of the $b'$ mass, $m_{b'} = 100$ GeV
we know
that $R_{CKM} \, < \, 8.4 \times 10^{-4}$ and so
\[
V_{cb'} \, < \, 8.4 \times 10^{-4} \sqrt{0.14/0.75} \,
\approx \,  3.6 \times 10^{-4} 
\]
with $m_{t'}= m_{b'} + 50  =  150$ GeV.
The bound gets weaker for smaller $m_{t'}$ \cite{Yanir:2002cq}.

Finally we mention that if new particles are added
and the $b'$ gets a new decay channel the excluded region
will always shrink. This can be seen in \cite{us}
where an exclusion plot with the Higgs channel
opened and a Higgs mass of 115 GeV is shown (see also
\cite{Arhrib:2000ct}).

\section{Conclusion}

Using all available
experimental data for  $m_{b'}> 96$ GeV we have
shown that there is still plenty of room for a
sequential $b'$
with a mass larger than 96 GeV.
This is our main conclusion.
We have also shown
that the allowed region depends heavily
on the values of $R_{CKM}$ and $m_{t'}$.
The region where $m_{t'} \approx m_{t}$
is always excluded due to a 100 \% CC. 
All plots show that $R_{CKM}$ is 
somewhere between $10^{-4}$
and  $10^{-2}$ which is a reasonable
value regarding the CKM elements
known so far.
In fact, those CKM values
suggest that $V_{cb'} \approx 10^{-4}-10^{-3}$. If 
$V_{tb'} \approx 10^{-1}$ then a value of 
$R_{CKM}$ between  $10^{-2}$ and $10^{-4}$ is 
absolutely natural. 
Moreover, the limit we have obtained for
$V_{cb'}$ in the last section makes it
even more natural.

We know that the DELPHI
and also the CDF analysis
are being improved and we hope
to shrink the allowed region with
this new data.

As for the future, searches in hadron colliders
will have to wait for the RunII of the Tevatron
and for the Large Hadron Collider (LHC). The $b' \bar{b'}$
production cross section increases by roughly two orders
of magnitude at the LHC compared to the Tevatron. Thus
LHC will be a copious source of $b'$ pairs. With high
values for cross section and luminosity, if background is suppressed
exclusion plots can be drawn for a very wide range of $b'$
masses.
For a detailed discussion on future searches see
\cite{Frampton:1999xi}. 

In summary we believe that there is still experimental
and theoretical work to be done to find or definitely
to exclude a sequential fourth generation of quarks
at the electroweak scale.

This work is supported
by Funda\c{c}\~ao para a Ci\^encia e Tecnologia under contract
POCTI/FNU/49523/2002. S.M.O. is supported by FCT
under contract SFRH/BD/6455/2001.

\end{document}